\documentclass{article}
\usepackage{amssymb}
\usepackage{amsmath}
\usepackage{graphicx}
\usepackage{epsfig}

\newtheorem{theorem}{Theorem}
\newtheorem{acknowledgement}[theorem]{Acknowledgement}

\input{tcilatex}

\begin{document}

\title{Quantum capacity of an amplitude-damping channel with memory}
\author{Rabia Jahangir$^{1}$, Nigum Arshed$^{2}$ and A. H. Toor$^{2}$ \\
%EndAName
$^{1}$\textit{National Centre for Physics, Shahdrah Valley Road}\\
\textit{Islamabad 44000 Pakistan}\\
$^{2}$\textit{Department of Physics, Quaid-i-Azam University}\\
\textit{Islamabad 45320, Pakistan}}
\maketitle

\begin{abstract}
We calculate the quantum capacity of an amplitude-damping channel with time
correlated Markov noise, for two channel uses. Our results show that memory
of the channel increases it's ability to transmit quantum information
significantly. We analyze and compare our findings with earlier numerical
results on amplitude-damping channel with memory. An upper bound on the
amount of quantum information transmitted over the channel in presence of
memory, for an arbitrary number of channel uses is also presented.
\end{abstract}

\section{Introduction}

The biggest constraint for reliable transmission of quantum information is
the presence of noise in quantum channels. It causes decoherence of quantum
systems resulting in the irreversible loss of information. The maximum
amount of information that can be reliably transmitted over a channel, per
channel use is known as it's capacity \cite{Shannon 1948,Thomas and Cover}.
Quantum channels, unlike their classical counterparts have more than one
capacities \cite{Bennett and Shor 2004}, depending on the type of
information transmitted, communication protocols and auxiliary resources
used. One of the fundamental tasks of quantum information theory is to
evaluate the capacities of quantum channels \cite{Bennett and Shor 1998}.

Early studies of quantum channel capacities mainly focused on memoryless
channels [5-20]. A channel is memoryless if it's action over each channel
use is independent of all other uses, that is, $\mathcal{E}_{N}=\mathcal{E}%
_{1}^{\otimes N}$. However, in real physical systems the channel action over
consecutive uses exhibits some correlation and $\mathcal{E}_{N}\neq \mathcal{%
E}_{1}^{\otimes N}$. These channels, known as quantum memory channels,
attracted lot of attention lately and a number of interesting results for
their capacities were reported [21-34].

The evaluation of quantum capacity for both memoryless and quantum memory
channels has been challenging. It is non-additive \cite{I Devetak 2005,
Barnum Nielson Schumacher 1998} therefore, the task to calculate it is not
trivial. Quantum capacity of memoryless degradable channels, for which
coherent information is additive \cite{Devetak and Shor 2005}, has been
determined \cite{Giovannetti and Fazio 2005,Wolf and Garcia 2007}. It was
also evaluated for a special class of memory channels known as forgetful
channels \cite{Benenti Arrigo and Falci 2009,Arrigo Benenti and Falci 2011}.
Recently, coding theorems for quantum capacity of long term memory channels
(not-forgetful) were proved [32-34].

In this paper we determine the quantum capacity $Q$ of an amplitude-damping
channel with finite memory. The noise over consecutive uses of the channel
is assumed to be time correlated Markov noise \cite{Macchiavello and Palma
2002}. We evaluate the quantum capacity analytically and give a comparison
with the numerical results reported in \cite{Arrigo Benenti and Falci 2011}.
Our results are in agreement with the findings of Ref. \cite{Arrigo Benenti
and Falci 2011} for the same set of channel parameters. We also calculate
quantum capacity of the amplitude-damping channel for the special case of
perfect memory and arbitrary number of channel uses. The quantum capacity
increases in the presence of memory converging to it's maximum value with
the number of channel uses.

The paper is organized as follows. In Section 2, we recall basic concepts of
quantum channels and quantum capacity. In Section 3 we discuss the model of
an amplitude damping channel with memory and briefly describe the double
blocking strategy and forgetful channels. We calculate the quantum capacity
of the amplitude damping channel with correlated noise, for two channel uses
in Section 4. In Section 5, the quantum capacity for an amplitude damping
channel with perfect memory is presented, for arbitrary number of channel
uses. Finally, in Section 6 we discuss our results and conclude.

\section{Quantum channel and quantum capacity}

Quantum channels model noise processes that occur in quantum systems due to
the interaction with their environment \cite{Nielson and Chuang}.
Mathematically, a quantum channel $\mathcal{E}$ is a completely positive and
trace preserving map of a quantum system from an initial state $\rho _{S}\in 
\mathcal{H}_{S}$ to the final state,%
\begin{equation}
\mathcal{E}\left( \rho _{S}\right) =\text{Tr}_{E}\left[ U\left( \rho
_{S}\otimes \left\vert e_{0}\right\rangle \left\langle e_{0}\right\vert
\right) U^{\dag }\right] ,  \label{Quantum Channel}
\end{equation}%
where $\rho _{E}=\left\vert e_{0}\right\rangle \left\langle e_{0}\right\vert
\in \mathcal{H}_{E}$ is the initial state of the environment. The system
undergoes a unitary evolution $U$ with it's environment and a partial trace
Tr$_{E}$ is performed over the environment to get the final state of the
quantum system. In Kraus representation \cite{Kraus 1983}, the action of a
quantum channel can be described as%
\begin{equation}
\mathcal{E}\left( \rho _{S}\right) =\sum_{k}A_{k}\rho _{S}A_{k}^{\dag },
\label{Kraus Representation}
\end{equation}%
and the Kraus operators $A_{k}=\left\langle e_{k}\right\vert \rho
_{E}\left\vert e_{0}\right\rangle $, satisfy the completeness relationship $%
\sum_{k}A_{k}^{\dag }A_{k}=I_{S}$. The conjugate $\widetilde{\mathcal{E}}$
of a quantum channel $\mathcal{E}$ is defined as \cite{King and Ruskai 2007},%
\begin{equation}
\widetilde{\mathcal{E}}\left( \rho _{S}\right) =\text{Tr}_{S}\left[ U\left(
\rho _{S}\otimes \left\vert e_{0}\right\rangle \left\langle e_{0}\right\vert
\right) U^{\dag }\right] .  \label{Conjugate Channel}
\end{equation}%
A quantum channel is degradable if it can be degraded to it's conjugate \cite%
{Devetak and Shor 2005}, that is, there exists a completely positive and
trace preserving map $\mathcal{T}$ such that $\widetilde{\mathcal{E}}=%
\mathcal{T}\circ \mathcal{E}$.

Quantum capacity $Q$ of a quantum channel depends on dimensions of the
largest subspace of $\mathcal{H}_{S}^{\otimes N}$\ that is reliably
transmitted over it, in the limit of large number of channel uses $N$ \cite%
{Seth Lloyd 1997,P W Shor 2002, I Devetak 2005}. For a memoryless channel,
it is given by the regularized coherent information $I_{c}$ \cite{Schumacher
and Westmoreland 1997}, maximized over all possible input states $\rho _{S}$
i.e.,%
\begin{equation}
Q=\lim_{N\rightarrow \infty }\frac{Q^{N}}{N},Q^{N}=\max_{\rho _{S}\in 
\mathcal{H}_{S}^{\otimes N}}I_{c}\left( \rho _{S},\mathcal{E}^{\otimes
N}\right) ,  \label{Quantum Capacity}
\end{equation}%
\begin{equation}
I_{c}=S\left[ \mathcal{E}^{\otimes N}\left( \rho _{S}\right) \right]
-S_{e}^{\otimes N}\left( \rho _{S}\right) .  \label{Coherent Information}
\end{equation}%
In Eq. (\ref{Coherent Information}), $S\left( \rho \right) =-$Tr$\left[ \rho
\log _{2}\rho \right] $ is the von Neumann entropy \cite{Nielson and Chuang}%
, and $S_{e}^{\otimes N}\left( \rho _{S}\right) $ is the entropy exchange
given by \cite{Schumacher 1996},%
\begin{equation}
S_{e}^{\otimes N}\left( \rho _{S}\right) =S\left[ \left( \mathcal{I}\otimes 
\mathcal{E}^{\otimes N}\right) \left( \left\vert \Psi \right\rangle
\left\langle \Psi \right\vert \right) \right] ,  \label{Entropy exchange-1}
\end{equation}%
where $\left\vert \Psi \right\rangle \in \mathcal{H}_{S}\otimes \mathcal{H}%
_{R}$ is a purification of $\rho _{S}$ obtained by appending a reference
Hilbert space $\mathcal{H}_{R}$ to the system Hilbert space $\mathcal{H}_{S}$%
. Entropy exchange can also be defined in terms of the conjugate channel as 
\cite{Schumacher 1996},%
\begin{equation}
S_{e}^{\otimes N}\left( \rho _{S}\right) =S\left[ \widetilde{\mathcal{E}}%
\left( \rho _{S}\right) \right] =S\left( \rho _{E}^{\prime }\right) .
\label{Entropy Exchange-2}
\end{equation}%
In the above expression, $\rho _{E}^{\prime }$ is the state of environment
after interaction with the quantum system and is written as 
\begin{equation}
\rho _{E}^{\prime }=\sum_{i,j}\text{Tr}_{S}\left( A_{i}\rho A_{j}^{\dag
}\right) \left\vert e_{i}\right\rangle \left\langle e_{j}\right\vert ,
\label{Environment Output}
\end{equation}%
where $\left\vert e_{i}\right\rangle $ are the orthonormal basis of the
environment $\mathcal{H}_{E}$. The coherent information $I_{c}$ is
superadditive \cite{Barnum Nielson Schumacher 1998}, therefore the limit $%
N\rightarrow \infty $ in Eq. (\ref{Quantum Capacity}) is necessary, which
makes the evaluation of $Q$ difficult. However, for degradable channels, $%
I_{c\text{ }}$ reduces to the conditional entropy which is subadditive and
concave \cite{Devetak and Shor 2005}. This is an important simplification as
for these channels the quantum capacity $Q$ is equal to the single shot
capacity $Q_{1}$.

\section{Amplitude-damping channel with memory}

Amplitude damping channel describes energy dissipation from a quantum system 
\cite{Nielson and Chuang}, such as, spontaneous emission of an atom and
relaxation of a spin system at high temperature into the equilibrium state.
It is a non-unital channel, i. e., $\mathcal{E}\left( I\right) \neq I$ \cite%
{King and Ruskai 2001}, with Kraus operators%
\begin{equation}
A_{0}=\left( 
\begin{array}{cc}
\cos \chi & 0 \\ 
0 & 1%
\end{array}%
\right) ,A_{1}=\left( 
\begin{array}{cc}
0 & 0 \\ 
\sin \chi & 0%
\end{array}%
\right) ,  \label{ADC Kraus}
\end{equation}%
where $\chi $ is the damping parameter with $0\leq \chi \leq \frac{\pi }{2}$%
. Memoryless amplitude damping channel is degradable \cite{Giovannetti and
Fazio 2005, Wolf and Garcia 2007}, therefore, it's quantum capacity can be
calculated using Eq. (\ref{Quantum Capacity}) and the single channel use
formula $Q=Q_{1}$ applies. However, when memory effects are taken into
account Eq. (\ref{Quantum Capacity}) only provides an upper bound on $Q$
which can be saturated for forgetful channels \cite{Kretschmann and Werner
2005}. Next we study an amplitude-damping channel with memory.

\subsection{The model}

\begin{figure}[tph]
\centering
\includegraphics[width=2.4in]{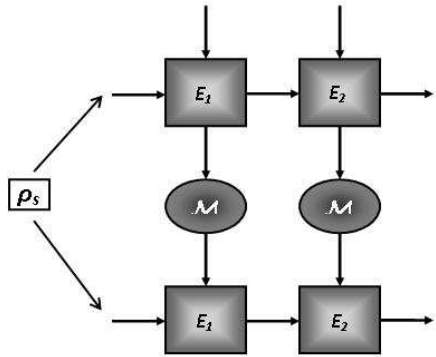}
\caption{Communication protocol}
\label{Communication Protocol}
\end{figure}

Consider an amplitude-damping channel with finite memory for two consecutive
uses, as shown in Fig. (\ref{Communication Protocol}). The time flows from
left to right while the horizontal arrows represent a two-qubit input state $%
\rho _{S}$. We assume that the Hilbert space of the environment $\mathcal{H}%
_{E}$, consists of two subspaces $\mathcal{H}_{E_{i}}$\ and $\mathcal{H}_{M}$
\cite{Bowen and Mancini 2004}. The finite subspace $\mathcal{H}_{M}$
represents\ memory of the channel which does not decay over the time scale
of consecutive channel uses. The subspace $\mathcal{H}_{E_{i}}$, where $%
i=1,2 $, represents two possible states of the environment which determine
the noise introduced by the channel given by Kraus operators in Eq. (\ref%
{ADC Kraus}). The vertical arrows in Fig. (\ref{Communication Protocol})
represent the influence of the interaction of first qubit with environment
on the second qubit.

Mathematically, amplitude-damping channel with time-correlated Markov noise 
\cite{Yeo and Skeen 2003}, is given by%
\begin{equation}
\mathcal{E}\left( \rho _{S}\right) =\left( 1-\mu \right)
\sum_{i,j=0}^{1}A_{ij}^{u}\rho _{S}A_{ij}^{u\dag }+\mu
\sum_{k=0}^{1}A_{kk}^{c}\rho _{S}A_{kk}^{c\dag },  \label{Partial ADC}
\end{equation}%
where $0\leq \mu \leq 1$, is the memory parameter. In the above expression
noise is uncorrelated with probability $\left( 1-\mu \right) $ and the
channel action is specified by Kraus operators%
\begin{equation}
A_{ij}^{u}=A_{i}\otimes A_{j},  \label{Un-correlated ADC}
\end{equation}%
where $i,j=0,1$ with $A_{i}$ given by Eq. (\ref{ADC Kraus}), while with
probability $\mu $\ it is correlated and specified by $A_{kk}^{c}$. The task
of constructing Kraus operators $A_{kk}^{c}$ for the non-Pauli
amplitude-damping channel is not trivial. In order to derive these operators
the Lindbladian is solved using damping basis \cite{Daffer Wodkiewicz McIver
2003, Yeo and Skeen 2003}. The map%
\begin{equation}
\Phi \left( \rho \right) =\exp \left( \mathcal{L}t\right) \rho =\sum_{i}%
\text{Tr}\left( L_{i}\rho \right) \exp \left( \lambda _{i}t\right) R_{i},
\label{Damping basis}
\end{equation}%
describes a wide class of Markov quantum channels, where $\lambda _{i}$ are
the damping eigenvalues. For a finite $N$-dimensional Hilbert space 
\begin{equation}
\mathcal{L}\rho =-\frac{1}{2}\sum_{i,j}^{N^{2}-1}c_{ij}\left( F_{j}^{\dag
}F_{i}\rho +\rho F_{j}^{\dag }F_{i}-2F_{i}\rho F_{j}^{\dag }\right) ,
\label{Lindbladian}
\end{equation}%
and the system operators satisfy Tr$(F_{i})=0$ and Tr$\left( F_{i}^{\dag
}F_{j}\right) =\delta _{ij}$ while $c_{ij}$ form a positive matrix. The
right eigenoperators $R_{i}$ satisfy the eigenvalue equation%
\begin{equation}
\mathcal{L}R_{i}=\lambda _{i}R_{i},  \label{Eigenvalue equation}
\end{equation}%
and the duality relation%
\begin{equation}
\text{Tr}\left( L_{i}R_{i}\right) =\delta _{ij},  \label{Duality relation}
\end{equation}%
with the left eigen operators $L_{i}$. The Lindbladian%
\begin{equation}
\mathcal{L}\rho _{S}=-\frac{1}{2}\alpha \left( \sigma ^{\dag }\sigma \rho
_{S}+\rho _{S}\sigma ^{\dag }\sigma -2\sigma \rho _{S}\sigma ^{\dag }\right)
,  \label{Lindbladian ADC}
\end{equation}%
yields the amplitude damping channel. The parameter $\alpha $ is analogous
to the Einstein coefficient of spontaneous emission \cite{Daffer Wodkiewicz
McIver 2003}, and $\sigma ^{\dag }\equiv \frac{1}{2}\left( \sigma
^{x}+i\sigma ^{y}\right) $ and $\sigma \equiv \frac{1}{2}\left( \sigma
^{x}-i\sigma ^{y}\right) $ are the raising and lowering operators,
respectively. In order to obtain time-correlated amplitude damping channel
for two channel uses, Eq. (\ref{Lindbladian ADC}) is modified to \cite{Yeo
and Skeen 2003}, 
\begin{eqnarray}
\mathcal{L}\rho _{S} &=&-\frac{\alpha }{2}\left[ \left( \sigma ^{\dag
}\otimes \sigma ^{\dag }\right) \left( \sigma \otimes \sigma \right) \rho
_{S}+\rho _{S}\left( \sigma ^{\dag }\otimes \sigma ^{\dag }\right) \left(
\sigma \otimes \sigma \right) \right.  \notag \\
&&\left. -2\left( \sigma \otimes \sigma \right) \rho _{S}\left( \sigma
^{\dag }\otimes \sigma ^{\dag }\right) \right] .
\label{Lindbladian ADC-Two uses}
\end{eqnarray}%
The resulting completely positive and trace preserving map is given by 
\begin{equation}
\mathcal{E}\left( \rho _{S}\right) =\sum_{i,j=0}^{3}\text{Tr}\left(
L_{ij}\rho _{S}\right) \exp \left( \lambda _{ij}t\right)
R_{ij}=\sum_{k=0}^{1}A_{kk}^{c}\rho _{S}A_{kk}^{c},  \label{Correlated ADC}
\end{equation}%
where the Kraus operators for correlated noise are%
\begin{equation}
A_{00}^{c}=\left( 
\begin{array}{cccc}
\cos \chi & 0 & 0 & 0 \\ 
0 & 1 & 0 & 0 \\ 
0 & 0 & 1 & 0 \\ 
0 & 0 & 0 & 1%
\end{array}%
\right) ,A_{11}^{c}=\left( 
\begin{array}{cccc}
0 & 0 & 0 & 0 \\ 
0 & 0 & 0 & 0 \\ 
0 & 0 & 0 & 0 \\ 
\sin \chi & 0 & 0 & 0%
\end{array}%
\right) ,  \label{Correlated ADC Kraus}
\end{equation}%
with $\cos \chi \equiv \exp \left( -\frac{1}{2}\alpha t\right) $ and $\sin
\chi \equiv \sqrt{1-\exp \left( -\alpha t\right) }$. The Eqs. (\ref{Partial
ADC}), (\ref{Un-correlated ADC}) and (\ref{Correlated ADC Kraus}) give the
amplitude damping channel with memory. In the presence of memory, amplitude
damping channel is not degradable, therefore, the maximization over the
input states $N\rightarrow \infty $, can not be avoided.

\subsection{Forgetful channels}

If the memory of a channel decays exponentially with time and it's output
depends weakly on the initialization of memory, it is known as a forgetful
channel \cite{Kretschmann and Werner 2005}. In order to obtain a forgetful
channel, double blocking strategy is used. Consider $N+L$ blocks of channel
uses. The actual coding and decoding is done for the first $N$ uses,
ignoring the remaining $L$ idle uses. The resulting CPT map $\overline{%
\mathcal{E}}_{N+L}$ acts on the states $\rho _{S}\in \mathcal{H}%
_{S}^{\otimes N}$. If we consider $M$ uses of such blocks, the corresponding
memory channel $\overline{\mathcal{E}}_{M\left( N+L\right) }$ can be
approximated by the memoryless channel $\left( \overline{\mathcal{E}}%
_{\left( N+L\right) }\right) ^{\otimes M}$. The correlations among different
blocks decay during the $L$ idle uses, which is expressed as%
\begin{equation}
\left\Vert \overline{\mathcal{E}}_{M\left( N+L\right) }\left( \rho \right)
-\left( \overline{\mathcal{E}}_{\left( N+L\right) }\right) ^{\otimes
M}\left( \rho \right) \right\Vert _{1}\leq h\left( m-1\right) c^{-L}
\label{Forgetfulness Condition}
\end{equation}%
where $\rho \in \mathcal{H}^{\otimes MN}$ is an input state, $h$ is a
constant depending on the memory model, $c>1$ ($c$ and $h$ are independent
to the input state) and $\left\Vert .\right\Vert _{1}$ is the trace norm.
The error committed by replacing the memory channel $\overline{\mathcal{E}}%
_{M\left( N+L\right) }$ with the corresponding memoryless channel $\left( 
\overline{\mathcal{E}}_{\left( N+L\right) }\right) ^{\otimes M}$, grows with
the number of blocks $M$, but it goes to zero exponentially fast with the
number $L$ of idle uses in a block. This maps a quantum memory channel into
memoryless one, with negligible error and permits the proof of coding
theorems for this class of channels \cite{Kretschmann and Werner 2005}.
Quantum capacity $Q$ for forgetful channels is given by%
\begin{equation}
Q=\lim_{N\rightarrow \infty }\frac{Q^{N}}{N},Q^{N}=\max_{\rho _{S}\in 
\mathcal{H}_{S}^{\otimes N}}I_{c}\left( \rho _{S},\mathcal{E}^{N}\right) ,
\label{Forgetful Quantum Capacity}
\end{equation}%
\begin{equation}
I_{c}=S\left[ \mathcal{E}^{N}\left( \rho _{S}\right) \right]
-S_{e}^{N}\left( \rho _{S}\right) .  \label{Forgetful Coherent Information}
\end{equation}

One might think that the capacity of $\mathcal{E}_{N+L}$ is greater than
that of $\overline{\mathcal{E}}_{\left( N+L\right) }$\ as the information
related to the idle uses $L$ is thrown away. However, the memory channel $%
\overline{\mathcal{E}}_{\left( N+L\right) }$ can faithfully transmit the
same amount of quantum information as the corresponding memoryless channel $%
\mathcal{E}_{N+L}$, with capacity given by Eq. (\ref{Quantum Capacity}).
There exists a simple proof (See Appendix A in Ref. \cite{Arrigo Benenti and
Falci 2011}) that%
\begin{equation}
I_{c}\left( \rho ^{\left( N+L\right) },\mathcal{E}_{N+L}\right) \leq
I_{c}\left( \rho ^{\left( N+L\right) },\overline{\mathcal{E}}_{N+L}\right)
+L\log _{2}\dim \left( \mathcal{H}_{S}\right) ,
\label{Coherent Information-Memory Channel}
\end{equation}%
therefore, quantum capacity of these two channels coincides for $%
\lim_{N\rightarrow \infty }\left( L/N\right) =0$.

\section{Quantum capacity of an amplitude-damping channel with memory}

We now calculate the quantum capacity $Q$ for the amplitude damping channel
with memory described above. Once again consider the communication system
shown in Fig. (\ref{Communication Protocol}). The information is encoded on
an input state $\rho _{S}^{(2)}$ which is a generic two-qubit state given by%
\begin{equation}
\rho _{S}^{(2)}=\rho _{S_{1}}\otimes \rho _{S_{2}},
\label{Two-qubit input state}
\end{equation}%
with%
\begin{equation}
\rho _{S_{i}}=\left( 1-p\right) \left\vert 0\right\rangle \left\langle
0\right\vert +r\left\vert 0\right\rangle \left\langle 1\right\vert +r^{\ast
}\left\vert 1\right\rangle \left\langle 0\right\vert +p\left\vert
1\right\rangle \left\langle 1\right\vert ,  \label{Single-qubit input state}
\end{equation}%
where $p$ is real and $\left\vert r\right\vert \leq \sqrt{p\left( 1-p\right) 
}$. The input $\rho _{S}$ is part of a larger system $SR$ which is in a pure
state $\left\vert \Psi \right\rangle $. The reference system $R$ does not
evolve and is a mathematical device to purify the $\rho _{S}$. The system $%
\rho _{S}=$Tr$_{R}\left( \left\vert \Psi \right\rangle \left\langle \Psi
\right\vert \right) $ couples with an environment $\rho _{E}$ undergoes a
unitary interaction. We know if a composite system $SR$, composed of two
subsystems $S$ and $R$, is in a pure state then $S\left( \rho _{S}\right)
=S\left( \rho _{R}\right) $ \cite{Nielson and Chuang}. In this case, neither
of the subsystems have any coherence. One might argue, what if the input $%
\rho _{S}$ is pure? The interaction between the system $\rho _{S}$ and
environment $\rho _{E}$ is unitary, therefore, the coherence lost by the
system is gained by the environment. If the input $\rho _{S}$ is pure, then
entropy of output state $\mathcal{E}\left( \rho _{S}\right) $\ and entropy
exchange $S_{e}\left( \rho _{S}\right) $ are equal. In this situation, the
coherent information $I_{c}$ and hence the quantum capacity $Q$\ is zero.
Therefore,coherent information $I_{c}$\ of an amplitude-damping channel is
maximized for an input with $r=0$ (for mathematical details see Appendix A),
therefore we set%
\begin{equation}
\rho _{S}^{(2)}=\left( 1-p\right) ^{2}\left\vert 00\right\rangle
\left\langle 00\right\vert +p\left( 1-p\right) \left( \left\vert
00\right\rangle \left\langle 11\right\vert +\left\vert 11\right\rangle
\left\langle 00\right\vert \right) +p^{2}\left\vert 11\right\rangle
\left\langle 11\right\vert .  \label{Input state}
\end{equation}%
This state is transmitted over the channel which maps it onto an output state%
\begin{equation}
\mathcal{E}\left( \rho _{S}^{(2)}\right) =\left( 1-\mu \right)
\sum_{i,j=0}^{1}A_{ij}^{u}\rho _{S}^{(2)}A_{ij}^{u\dag }+\mu
\sum_{k=0}^{1}A_{kk}^{c}\rho _{S}^{(2)}A_{kk}^{c\dag },  \label{Output state}
\end{equation}%
with eigenvalues%
\begin{eqnarray}
\lambda _{1} &=&p^{2}\cos ^{2}\chi \left( \cos ^{2}\chi +\mu \sin ^{2}\chi
\right) ,  \notag \\
\lambda _{2} &=&\lambda _{3}=p\cos ^{2}\chi \left( 1-p\cos ^{2}\chi \right)
+\mu p\sin ^{2}\chi \left[ 1-p\left( 1+\cos ^{2}\chi \right) \right] , 
\notag \\
\lambda _{4} &=&\left( 1-p\cos ^{2}\chi \right) ^{2}+\mu p\sin ^{2}\chi 
\left[ p\cos ^{2}\chi -2\left( 1-p\right) \right] .
\label{Output state eigenvalues}
\end{eqnarray}%
We assume without loss of generality that initially the state of the
environment is pure, 
\begin{equation}
\rho _{E}=\left( \left\vert 00\right\rangle \left\langle 00\right\vert
\right) _{E},  \label{Environment input}
\end{equation}%
which after interaction with the input state $\rho _{S}^{(2)}$ is modified to%
\begin{eqnarray}
\rho _{E}^{\prime } &=&\left( 1-\mu \right) \sum_{i,j,k,l=0}^{1}\text{Tr}%
_{S}\left( A_{ij}^{u}\rho _{S}^{(2)}A_{kl}^{u\dag }\right) \left\vert
e_{ij}\right\rangle \left\langle e_{kl}\right\vert  \notag \\
&&+\mu \sum_{m,n=0}^{1}\text{Tr}_{S}\left( A_{mm}^{c}\rho
_{S}^{(2)}A_{nn}^{c\dag }\right) \left\vert e_{mm}\right\rangle \left\langle
e_{nn}\right\vert ,  \label{Environment Output-2}
\end{eqnarray}%
where $\left\vert e_{ij}\right\rangle =\left\vert e_{i}\right\rangle \otimes
\left\vert e_{j}\right\rangle $ are the orthonormal basis of the
environment. The eigenvalues of the output state $\rho _{E}^{\prime }$ are%
\begin{eqnarray}
\widetilde{\lambda }_{1} &=&p^{2}\sin ^{2}\chi \left( \sin ^{2}\chi +\mu
\cos ^{2}\chi \right) ,  \notag \\
\widetilde{\lambda }_{2} &=&\widetilde{\lambda }_{3}=\left( 1-\mu \right)
p\sin ^{2}\chi \left( 1-p\sin ^{2}\chi \right) ,  \notag \\
\widetilde{\lambda }_{4} &=&\left( 1-p\sin ^{2}\chi \right) ^{2}+\mu p\sin
^{2}\chi \left[ 2-p\left( 1+\sin ^{2}\chi \right) \right] .
\label{Environment Output-2 Eigenvalues}
\end{eqnarray}%
The quantum capacity of the amplitude damping channel with Markov correlated
noise, for two channel uses is calculated using Eqs. (\ref{Forgetful Quantum
Capacity}) and (\ref{Forgetful Coherent Information}) which gives%
\begin{equation}
Q^{2}=-\max_{p}\sum_{i=1}^{4}\left( \lambda _{i}\log _{2}\lambda _{i}-%
\widetilde{\lambda }_{i}\log _{2}\widetilde{\lambda }_{i}\right) .
\label{Quantum Capacity 2 uses}
\end{equation}%
The maximization over the input probability $p$ is performed numerically. If
the channel is noiseless i.e., $\chi =0\ $then the capacity $Q^{2}$ is
maximized for a maximally mixed input state with $p=\frac{1}{2}$. However,
our results show that for $\chi \neq 0$ it is not the optimal choice for the
input state. This is consistent with the earlier work for quantum capacity
with correlated noise \cite{Benenti Arrigo and Falci 2009, Arrigo Benenti
and Falci 2011}.

\begin{figure}[tph]
\centering
\includegraphics[width=2.4in]{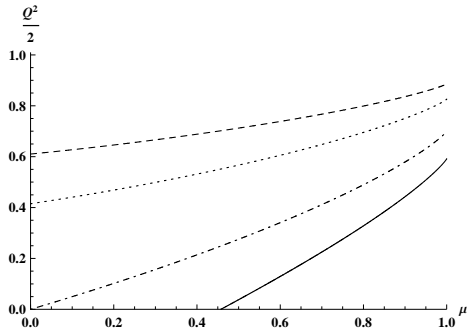}
\caption{Quantum capacity $\frac{Q^{2}}{2}$ as a function of the memory
parameter $\protect\mu $ for $\protect\chi =\frac{\protect\pi }{8}$, $%
p=0.465 $ (dashed), $\protect\chi =\frac{\protect\pi }{6}$, $p=0.447$
(dotted), $\protect\chi =\frac{\protect\pi }{4}$, $p=0.389$ (dotted-dashed)
and $\protect\chi =\frac{\protect\pi }{3}$, $p=0.312$ (solid). The capacity
is normalized with respect to the number of channel uses.}
\label{Q_2 versus Meu}
\end{figure}

In Fig. (\ref{Q_2 versus Meu}) we plot the quantum capacity $\frac{Q^{2}}{2}$
of the amplitude damping channel with correlated noise (normalized with
respect to the number of channel uses) versus the memory coefficient $\mu $,
for different values of the damping parameter $\chi $. Our results show that
when the channel noise increases, the input state maximizing Eq. (\ref%
{Quantum Capacity 2 uses}) becomes less than maximally mixed. It is evident
from the plots that as the channel changes from memoryless $\mu =0$ to a
perfect memory channel $\mu =1$, the quantum capacity increases, for all
values of $\chi $. In particular note that for $\chi =\frac{\pi }{4}$, the
capacity is zero but if the channel has non-zero memory it can transmit
quantum information. Similarly, for $\chi =\frac{\pi }{3}$ the quantum
capacity of the amplitude damping channel is zero unless the memory
coefficient $\mu \gtrsim \frac{1}{2}$, beyond this value of $\mu $\ the
increase in $Q^{2}$ is almost linear. We infer that memory of an amplitude
damping channel increases it's capacity to transmit quantum information
significantly.

\subsection{Example: Damped harmonic oscillator}

An interesting example of amplitude-damping channel with memory is given by
a damped harmonic oscillator \cite{Arrigo Benenti and Falci 2011}. In this
model, a stream of $N$ qubits (the system $Q$) interacts with the local
environment of a harmonic oscillator $O$\ coupled with a reservoir. The
Hamiltonian of the total system is%
\begin{equation}
\mathcal{H}\left( t\right) =\mathcal{H}_{0}+\mathcal{H}_{QO}+\delta \mathcal{%
H},  \label{Total Hamiltonian}
\end{equation}%
with%
\begin{equation}
\mathcal{H}_{0}=\mathcal{H}_{Q}+\mathcal{H}_{O}=-\frac{\omega }{2}%
\sum_{k=1}^{N}\sigma _{k}^{z}+\upsilon \left( a^{\dag }a+\frac{1}{2}\right) ,
\label{Unitary Hamiltonian}
\end{equation}%
and%
\begin{equation}
\mathcal{H}_{QO}=\lambda \sum_{k=1}^{N}f_{k}\left( t\right) \left( a^{\dag
}\sigma _{k}+a\sigma _{k}^{\dag }\right) .  \label{Interaction Hamiltonian}
\end{equation}%
Here $\sigma _{z}=\left\vert g\right\rangle \left\langle g\right\vert
+\left\vert e\right\rangle \left\langle e\right\vert $, while $a$ and $%
a^{\dag }$ are the creation and annihilation operators of the harmonic
oscillator. The interaction $\mathcal{H}_{QO}$ between the qubit and
harmonic oscillator is of Jaynes-Cummings type. The constant $\lambda $ is
real and positive and $\hbar =1$. The qubit-harmonic oscillator coupling $%
f_{k}\left( t\right) =1$, when qubit $k$ is inside the channel while $%
f_{k}\left( t\right) =0$\ otherwise. Two consecutive qubits entering the
channel are separated by a time interval $\tau $, and the qubit transit time
is $\tau _{p}$. The term $\delta \mathcal{H}$ describes the reservoir
Hamiltonian and the local environment-reservoir interaction which damps the
oscillator within dissipation time scale $\tau _{d}$.

The oscillator damping is described by%
\begin{equation}
\frac{d}{dt}\rho _{O}=\Gamma \left( a\rho _{O}a^{\dag }-\frac{1}{2}a^{\dag
}a\rho _{O}-\frac{1}{2}\rho _{O}a^{\dag }a\right) ,
\label{Oscillator master equation}
\end{equation}%
which asymptotically decays to the ground state $\left\vert 0\right\rangle $
with rate $\Gamma $\ and $\tau _{d}=\frac{1}{\Gamma }$. In this case, the
memory parameter can be defined as%
\begin{equation}
\mu \equiv \frac{\tau _{d}}{\tau +\tau _{d}}.  \label{Memory parameter}
\end{equation}%
The memoryless limit $\mu \ll 1$ is achieved for fast decay $\tau _{d}\ll
\tau $ whereas for $\tau _{d}\approx \tau $, the memory parameter $\mu
\lesssim 1$ and memory effects come into play. Arrigo \textit{et. al.,}
solved this system numerically for two channel uses \cite{Arrigo Benenti and
Falci 2011}.

\begin{figure}[h]
\centering
\includegraphics[width=3.4in]{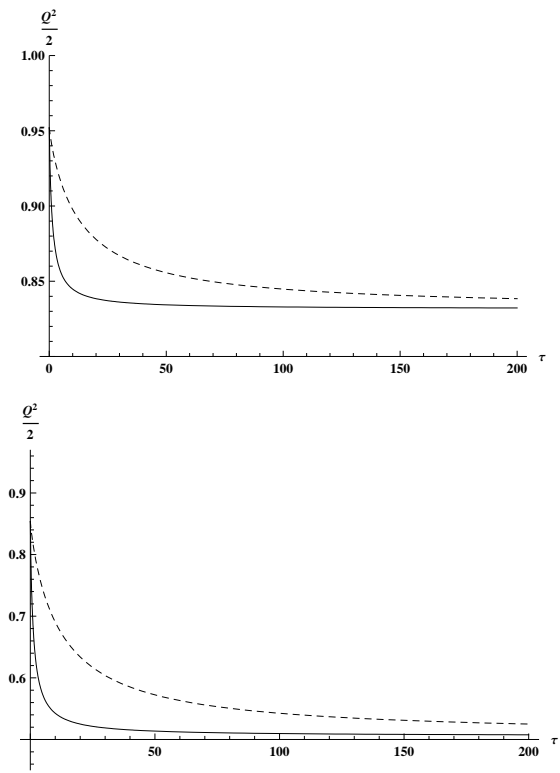}
\caption{Capacity $\frac{Q^{2}}{2}$ as a function of the time inteval
between two consecutive channel uses $\protect\tau $ with $\protect\chi %
=0.225$, $p=0.486$ (top) and $\protect\chi =0.464$, $p=0.456$ (bottom) for $%
\protect\tau _{d}=20$ (dashed) and $\protect\tau _{d}=2$ (solid). See text
for additional details.}
\label{Q_2 versus Tau}
\end{figure}

In Fig. (\ref{Q_2 versus Tau}), we plot $\frac{Q^{2}}{2}$ versus $\tau $ for
the same values of channel parameters as in Ref. \cite{Arrigo Benenti and
Falci 2011}. In Eq. (\ref{Quantum Capacity 2 uses}) we redefine $\mu $ as
given by Eq. (\ref{Memory parameter}) and consider both weak $\tau _{d}=20$
and strong $\tau _{d}=2$ dissipation strength $\Gamma =\frac{1}{\tau _{d}}$
of the oscillator. The memoryless limit is recovered for $\tau \rightarrow
\infty $ (for which the memory parameter $\mu \equiv \tau _{d}/\left( \tau
+\tau _{d}\right) \rightarrow 0$). When the channel is less noisy with the
damping parameter $\chi =0.225$, $\frac{Q^{2}}{2}$ is maximized for an input
state with $p=0.486$. In this case, when $\tau _{d}$ is large, the
oscillator damps to it's ground state slowly and the channel retains it's
memory for a longer period of time. Therefore, the amount of quantum
information transmitted over the channel is higher which decays to it's
minimum value at a relatively slower rate as we increase the time interval $%
\tau $ between two successive qubits entering the channel. In comparison,
when $\tau _{d}$ is small the decay rate of the oscillator $\Gamma $ is
large and it's state decays before the second qubit enters the channel. The
resulting channel is memoryless. It is evident from the plot that the
quantum capacity rapidly decreases to it's minimum value in this case. We
also plot $\frac{Q^{2}}{2}$ for a lower quality channel with $\chi =0.464$
and an input state with $p=0.456$. As the channel is more noisy than the
previous case therefore, less amount of quantum information is transmitted.
However, once again we witness larger capacity and relatively gradual decay
to it's minimum for $\tau _{d}=20$, while smaller capacity and rapid decay
for $\tau _{d}=2$. These results are consistent with the findings of Ref. 
\cite{Arrigo Benenti and Falci 2011}.

\section{Quantum capacity of an amplitude-damping channel with perfect memory%
}

Finally, we consider a special case of sending an arbitrary number of qubits 
$N$ through an amplitude damping channel with perfect memory. It is assumed
that the state of the environment remains intact for the $N$ uses. This
gives an upper bound on the quantum capacity with correlated noise. Once
again the double block strategy is used to make sure that the channel action
over different blocks is not correlated.

The state input to the channel is an $N$ qubit state%
\begin{equation}
\rho _{S}^{N}=\rho _{S_{1}}\otimes \ldots \otimes \rho _{S_{N}},
\label{N qubit input state}
\end{equation}%
with $\rho _{S_{i}}=\left( 1-p\right) \left\vert 0\right\rangle \left\langle
0\right\vert +p\left\vert 1\right\rangle \left\langle 1\right\vert $. This
state is transmitted over an amplitude damping channel over $N$ uses of the
channel. If the noise over all successive uses of the channel is perfectly
uncorrelated then it maps the input state to%
\begin{equation}
\mathcal{E}\left( \rho _{S}^{N}\right) =\sum_{u_{1}\ldots u_{N}}\left(
A^{u_{1}}\otimes \ldots \otimes A^{u_{N}}\right) \rho _{S}^{N}\left(
A^{u_{1}\dag }\otimes \ldots \otimes A^{u_{N}\dag }\right) ,
\label{N qubit output state}
\end{equation}%
where $A^{u_{i}}\in \left[ A_{o},A_{1}\right] $ and $A_{i}$ with $i=0,1$ are
given in Eq. (\ref{ADC Kraus}). The eigenvalues of the output state are%
\begin{equation}
\gamma _{X}=p^{N}\left( \cos ^{2}\chi \right) ^{(N-X)}\left( 1-p\cos
^{2}\chi \right) ^{X},  \label{N qubit output eigenvalues}
\end{equation}%
with $X=0,\ldots ,N$. During the transmission, the input state is coupled
with an environment initially assumed to be in a pure state%
\begin{equation}
\rho _{E}^{N}=\left( \left\vert 0\right\rangle \left\langle 0\right\vert
\right) _{E}^{\otimes N}.  \label{N qubit environment input state}
\end{equation}%
As a result of the interaction with the input state $\rho _{S}^{N}$, it
modifies to an output state $\rho _{E}^{\prime N}$ with eigenvalues 
\begin{equation}
\widetilde{\gamma }_{X}=p^{N}\left( \sin ^{2}\chi \right) ^{(N-X)}\left(
1-p\sin ^{2}\chi \right) ^{X},
\label{N qubit environment output eigenvalues}
\end{equation}%
where $X=0,\ldots ,N$. The quantum capacity can be calculated using Eqs. (%
\ref{Forgetful Quantum Capacity}) and (\ref{Forgetful Coherent Information})
which gives%
\begin{equation}
Q^{N}=-\underset{p}{\max }\sum_{x=0}^{N}\left( 
\begin{array}{c}
N \\ 
X%
\end{array}%
\right) \left[ \gamma _{X}\log _{2}\gamma _{X}-\widetilde{\gamma }_{X}\log
_{2}\widetilde{\gamma }_{X}\right] ,
\label{Quantum capacity N uses uncorrelated}
\end{equation}%
here the binomial $\left( 
\begin{array}{c}
N \\ 
X%
\end{array}%
\right) $ takes into account the number of times a particular eigenvalue is
repeated.

Next we consider the case when the noise over the successive uses of the
channel is perfectly correlated. In order to determine the Kraus operators
we solve the $N$ dimensional Lindbladian%
\begin{equation}
\mathcal{L}\rho _{S}^{N}=-\frac{1}{2}\alpha \left( \left( \sigma ^{\dag
}\right) ^{\otimes N}\left( \sigma \right) ^{\otimes N}\rho _{S}^{N}+\rho
_{S}^{N}\left( \sigma ^{\dag }\right) ^{\otimes N}\left( \sigma \right)
^{\otimes N}-2\left( \sigma \right) ^{\otimes N}\rho _{S}^{N}\left( \sigma
^{\dag }\right) ^{\otimes N}\right) ,  \label{N dimensional Lindbladian}
\end{equation}%
using the damping basis method outlined in Section 3. The resulting Kraus
operator are%
\begin{eqnarray}
A_{0}^{c_{N}} &=&\left( 
\begin{array}{ccccc}
\cos \chi & 0 & \cdots & 0 & 0 \\ 
0 & 1 & \cdots & 0 & 0 \\ 
\vdots & \vdots & \ddots & \vdots & \vdots \\ 
0 & 0 & \cdots & 1 & 0 \\ 
0 & 0 & \cdots & 0 & 1%
\end{array}%
\right) _{2^{N}\times 2^{N}},  \notag \\
A_{1}^{c_{N}} &=&\left( 
\begin{array}{ccccc}
0 & 0 & \cdots & 0 & 0 \\ 
0 & 0 & \cdots & 0 & 0 \\ 
\vdots & \vdots & \ddots & \vdots & \vdots \\ 
0 & 0 & \cdots & 0 & 0 \\ 
\sin \chi & 0 & \cdots & 0 & 0%
\end{array}%
\right) _{2^{N}\times 2^{N}}.  \label{Perfect memory Kraus operators N uses}
\end{eqnarray}%
If the input state $\rho _{S}^{N}$ given by Eq. (\ref{N qubit input state})
is transmitted over the amplitude damping channel with perfect memory then
it is mapped to an output state%
\begin{equation}
\mathcal{E}\left( \rho _{S}^{N}\right) =\sum_{k=0}^{1}A_{k}^{c_{N}}\rho
_{S}^{N}A_{k}^{\dag c_{N}},  \label{N dimensional perfect memory output}
\end{equation}%
with eigenvalues%
\begin{eqnarray}
\lambda _{1} &=&p^{N}\cos ^{2}\chi ,  \notag \\
\lambda _{2} &=&\left( 1-p\right) ^{N}+p^{N}\sin ^{2}\chi ,  \notag \\
\lambda ^{Y} &=&p^{N-Y}\left( 1-p\right) ^{Y},
\label{N dimensional perfect memory output eigenvalues}
\end{eqnarray}%
where $Y=1,\ldots ,N-1$. Similarly, the initial state of the environment $%
\rho _{E}^{N}$ is mapped to a state with two non-zero eigenvalues%
\begin{eqnarray}
\widetilde{\lambda }_{1} &=&p^{N}\sin ^{2}\chi ,  \notag \\
\widetilde{\lambda }_{2} &=&1-p^{N}\sin ^{2}\chi .
\label{Environment perfect memory output eigenvalues}
\end{eqnarray}%
Using Eqs. (\ref{Forgetful Quantum Capacity}) and (\ref{Forgetful Coherent
Information}) the quantum capacity is given by%
\begin{equation}
Q^{N}=-\underset{p}{\max }\left[ \sum_{i=1}^{2}\left( \lambda _{i}\log
_{2}\lambda _{i}-\widetilde{\lambda }_{i}\log _{2}\widetilde{\lambda }%
_{i}\right) +\sum_{Y=1}^{N-1}\left( 
\begin{array}{c}
N \\ 
Y%
\end{array}%
\right) \lambda ^{Y}\log _{2}\lambda ^{Y}\right] .
\label{Quantum capacity N uses perfect memory}
\end{equation}%
The maximization over the input probability $p$ for the quantum capacity
with memoryless and perfect memory channel given by Eqs. (\ref{Quantum
capacity N uses uncorrelated}) and (\ref{Quantum capacity N uses perfect
memory}), respectively is performed numerically. Our calculations show that
as the number of channel uses increases, the input state maximizing the
capacity $Q^{N}$ changes from a less than maximally mixed state to maximally
mixed state. We must evaluate the quantum capacity in the limit $%
N\rightarrow \infty $, therefore we infer that the maximally mixed state
maximizes $Q^{N}$.

\begin{figure}[tph]
\centering
\includegraphics[width=2.4in]{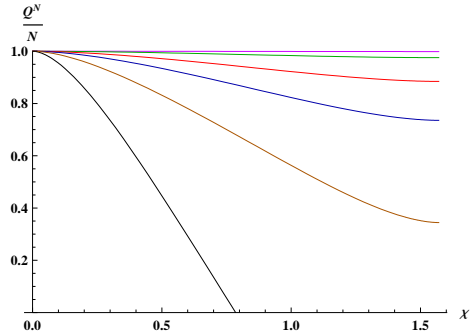}
\caption{Capacity $\frac{Q^{N}}{N}$ versus the channel noise $\protect\chi $
for memoryless amplitude damping channel $\protect\mu =0$ (black) and with
perfect memory $\protect\mu =1$ for $N=2$ (brown), $N=3$ (blue), $N=4$
(red), $N=6$ (green) and $N=10$ (purple).}
\label{Q_N versus Kie}
\end{figure}

In Fig. (\ref{Q_N versus Kie}) we plot $\frac{Q^{N}}{N}$ for both memoryless
and perfect memory amplitude damping channel, versus channel noise parameter 
$\chi $. We set $p=\frac{1}{2}$, for all channel uses. The memoryless
amplitude damping channel is degradable and it's quantum capacity $Q$ is
additive given by $Q=Q_{1}$, hence we have only one curve in the plot for
this case. It shows that the quantum capacity decreases to zero as the
memoryless amplitude damping channel becomes noisy. However, in the presence
of perfect memory we will always have non-zero quantum capacity, even if the
channel noise is maximum i.e., $\chi =\frac{\pi }{2}$. It is evident that as
the number of channel uses increase, the capacity converges to it's maximum
value $Q=1$.

\section{Conclusion}

We have studied an amplitude-damping with memory and calculated it's
capacity to transmit quantum information. Our results show that the quantum
capacity of the channel increases if the noise over the consecutive uses of
the channel is correlated, independent of the damping parameter $\chi $. If
the memory of the channel $\mu \gtrsim \frac{1}{2}$, we will always have
non-zero quantum capacity even if the channel noise attains it's maximum
allowed value. As compared to the numerical results reported in \cite{Arrigo
Benenti and Falci 2011} for an amplitude-damping channel with memory we
determine the quantum capacity analytically. The comparison of our study
with their results for the same set of channel parameter shows similar
behaviour with the increase of channel memory. In the case of
amplitude-damping channel with perfect memory channel, we calculate the
quantum capacity, for an arbitrary number of channel uses. Our results show
that the quantum capacity will always be non-zero if the channel has perfect
memory and saturates to it's maximum value in the limit $N\rightarrow \infty 
$.

\appendix{}

\section{Dependence of coherent information on $\left\vert r\right\vert ^{2}$%
}

Consider the two-qubit input state $\rho _{S}^{(2)}$ given by Eq. (\ref%
{Two-qubit input state}). For mathematical simplicity, we diagonalize the
input state such that%
\begin{equation}
\rho _{S_{i}}=\frac{1}{2}\left[ \left( 1-\Gamma \right) \left\vert
0\right\rangle \left\langle 0\right\vert +\left( 1+\Gamma \right) \left\vert
1\right\rangle \left\langle 1\right\vert \right] ,
\label{Single qubit diagonalized input}
\end{equation}%
where $\Gamma =\sqrt{(1-2p)^{2}+4\left\vert r\right\vert ^{2}}$\ and $0\leq
\left\vert r\right\vert ^{2}\leq p\left( 1-p\right) $. This state is
transmitted over an amplitude-damping channel with memory given by Eq. (\ref%
{Partial ADC}), which maps it to an output state with eigenvalues%
\begin{eqnarray}
\upsilon _{1} &=&\frac{1}{4}\left( 1-\Gamma \right) ^{2}\cos ^{2}\chi \left(
\cos ^{2}\chi +\mu \sin ^{2}\chi \right) ,  \notag \\
\upsilon _{2} &=&\upsilon _{3}=\frac{1}{2}\cos ^{2}\chi \left[ \left(
1-\Gamma \right) \sin ^{2}\chi +2\Lambda \cos ^{2}\chi \right]  \notag \\
&&+\mu \left[ \Lambda \left( 1-\cos ^{4}\chi \right) -\frac{1}{2}\left(
1-\Gamma \right) \cos ^{2}\chi \sin ^{2}\chi \right] ,  \notag \\
\upsilon _{4} &=&\frac{1}{2}\left[ 1+\sin ^{4}\chi +\Gamma \left( 1-\sin
^{4}\chi \right) -2\Lambda \cos ^{4}\chi \right.  \notag \\
&&\left. +\mu \left( 1-\Gamma \right) \cos ^{2}\chi \sin ^{2}\chi -2\Lambda
\sin ^{2}\chi \left( 2+\cos ^{2}\chi \right) \right] ,
\label{Output state for r}
\end{eqnarray}%
with $\Lambda =p\left( 1-p\right) -\left\vert r\right\vert ^{2}$. We assume
that the environment is initially in a pure state $\rho _{E}$ given by Eq. (%
\ref{Environment input}). The interaction with $\rho _{S}^{(2)}$, modifies
the environment to an output state $\rho _{E}^{\prime }$ with eigenvalues%
\begin{eqnarray}
\widetilde{\upsilon }_{1} &=&\frac{1}{4}\left( 1-\Gamma \right) ^{2}\sin
^{2}\chi \left( \sin ^{2}\chi +\mu \cos ^{2}\chi \right) ,  \notag \\
\widetilde{\upsilon }_{2} &=&\widetilde{\upsilon }_{3}=\frac{1}{4}\left(
1-\mu \right) \sin ^{2}\chi \left( 4\Lambda +\left( 1-\Gamma \right)
^{2}\cos ^{2}\chi \right) ,  \notag \\
\widetilde{\upsilon }_{4} &=&\frac{1}{4}\left[ 4\Gamma +\left( 1-\Gamma
\right) ^{2}\left( 1+\cos ^{4}\chi \right) +8\Lambda \cos ^{2}\chi +\mu
\left\{ 4\Lambda \left( \sin ^{2}\chi -2\cos ^{2}\chi \right) \right. \right.
\notag \\
&&\left. \left. +2\left( 1-\Gamma \right) \left( 1+\cos ^{2}\chi \right)
-\left( 1-\Gamma \right) ^{2}\left( 1+\cos ^{4}\chi \right) \right\} \right]
.  \label{Environment output r}
\end{eqnarray}%
Here, we have used Eq. (\ref{Environment Output-2}), to determine $\rho
_{E}^{\prime }$ with $\rho _{S_{i}}$ given by Eq. (\ref{Single qubit
diagonalized input}). Once we have the eigenvalues for the output states,
the coherent information $I_{c}$ and quantum capacity $\frac{Q^{2}}{2}$, for
two channel uses, is calculated from Eqs. (\ref{Forgetful Quantum Capacity})
and (\ref{Forgetful Coherent Information}), respectively.

\begin{figure}[h]
\centering
\includegraphics[width=2.4in]{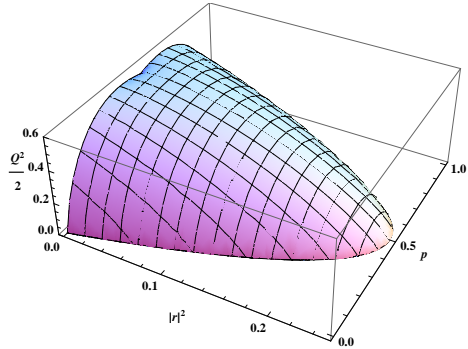}
\caption{Capacity $\frac{Q^{2}}{2}$ versus the input probability
distribution $p$ and coherence \ $\left\vert r\right\vert ^{2}$, for $%
\protect\mu =0.8$ and channel parameter $\protect\chi =0.685$.}
\label{Q_2 versus Modulus r and p}
\end{figure}

In Fig. (\ref{Q_2 versus Modulus r and p}), we plot $\frac{Q^{2}}{2}$ versus
the probability of input state $p$ and coherence $\left\vert r\right\vert
^{2}$, for fixed values of channel noise $\chi $ and memory $\mu $. It shows
that for all possible values of $p$, the capacity\ decreases as we increase
coherence, acquiring it's minimum value zero when coherence of the input
state is maximum, that is, $\left\vert r\right\vert ^{2}=p\left( 1-p\right) $%
. We conclude that coherent information $I_{c}$, is maximized for an input
state for which $\left\vert r\right\vert ^{2}=0$. The maximization over $p$,
in Eq. (\ref{Forgetful Quantum Capacity}) is performed numerically. For an
amplitude-damping channel with $\chi =0.685$, $\frac{Q^{2}}{2}$ is maximized
for $p=0.4154$. Therefore, in Fig. (\ref{Q_2 versus Modulus r and p}), there
is a slight decrease in $\frac{Q^{2}}{2}$, for $\left\vert r\right\vert
^{2}=0$ at $p=0.5$. If $\left\vert r\right\vert ^{2}=0$ then $\Gamma =1-2p$
and $\Lambda =p\left( 1-p\right) $, the eigenvalues of Eq. (\ref{Output
state eigenvalues}) and Eq. (\ref{Environment Output-2 Eigenvalues}) are
retrieved and the results are discussed in Section 4 . However, when the
coherence is maximum, then $\Gamma =1$ and $\Lambda =0$, both the system and
environment are pure and quantum capacity $\frac{Q^{2}}{2}$ is zero. This is
in agreement with the findings of Refs. \cite{Giovannetti and Fazio 2005,
Arrigo Benenti and Falci 2011}.

\begin{acknowledgement}
N. Arshed was supported by the Higher Education Commission Pakistan under
Grant No. 063-111368-Ps3-001.
\end{acknowledgement}

\end{document}